\PassOptionsToPackage{hyphens}{url}
\documentclass[letterpaper,twocolumn,10pt]{article}
\usepackage{usenix2019,epsfig,endnotes}

\usepackage{adjustbox}
\usepackage{enumerate}
\usepackage{graphicx}
\usepackage{epstopdf}
\usepackage{epsfig}
\usepackage{color}
\usepackage{url}
\usepackage{etoolbox}
\usepackage{booktabs}
\usepackage{amsmath}
\usepackage{amssymb}
\usepackage{mathtools}
\usepackage{hhline}
\usepackage{setspace}
\usepackage{balance}
\usepackage{caption}
\usepackage{subcaption}
\usepackage{xcolor,colortbl}
\usepackage{array}
\usepackage[ruled, noend]{algorithm2e}

\usepackage{balance}

\usepackage{multirow}
\usepackage{mathrsfs}
\usepackage{tabularx}
\usepackage{cleveref}
\usepackage{spverbatim}
\usepackage{float}
\usepackage[multiple]{footmisc}

\crefname{section}{\S}{\S}

\definecolor{LightCyan}{rgb}{0.88,1,1}
\definecolor{Gray}{gray}{0.9}

\newcommand{\fw}{Talon}
\newcommand{\name}{CDT}
\newcommand{\lname}{cross-device tracking\xspace}

\newcolumntype{x}[1]{>{\centering\arraybackslash\hspace{0pt}}p{#1}}

\newenvironment {squishlist}
{\begin{list}{$\bullet$}
  { \setlength{\itemsep}{1pt}
     \setlength{\parsep}{1pt}
     \setlength{\topsep}{1pt}
     \setlength{\partopsep}{1pt}
     \setlength{\leftmargin}{1.5em}
     \setlength{\labelwidth}{1em}
     \setlength{\labelsep}{0.5em} } }
{\end{list}}

\begin{document}

\date{}

\title{\Large \bf Talon: An Automated Framework for Cross-Device Tracking Detection}

\author{
\makebox[.4\linewidth]{\rm Konstantinos Solomos}\\
	FORTH, Greece\\
	solomos@ics.forth.gr\\
\and \makebox[.4\linewidth]{\rm Panagiotis Ilia}\\
	Univ. of Illinois at Chicago, USA\\
	pilia@uic.edu\\
\and \makebox[.4\linewidth]{\rm Sotiris Ioannidis}\\
	FORTH, Greece\\
	sotiris@ics.forth.gr\\
\and \makebox[.4\linewidth]{\rm Nicolas Kourtellis}\\
	Telefonica Reasearch, Spain\\
	nicolas.kourtellis@telefonica.com\\
} 

	
	\maketitle

\thispagestyle{empty}

	\subsection*{Abstract}
Although digital advertising fuels much of today's free Web, it typically does so at the cost of online users' privacy, due to the continuous tracking and leakage of users' personal data.
In search for new ways to optimize the effectiveness of ads, advertisers have introduced new advanced paradigms such as cross-device tracking (\name), to monitor users' browsing on multiple devices and screens, and deliver (re)targeted ads in the most appropriate screen.
Unfortunately, this practice leads to greater privacy concerns for the end-user.

Going beyond the state-of-the-art, we propose a novel methodology for detecting \name\ and measuring the factors affecting its performance, in a repeatable and systematic way.
This new methodology is based on emulating realistic browsing activity of end-users, from different devices, and thus triggering and detecting cross-device targeted ads.
We design and build \fw\footnote{https://en.wikipedia.org/wiki/Talos}, a \name\ measurement framework that implements our methodology and allows experimentation with multiple parallel devices, experimental setups and settings.
By employing \fw, we perform several critical experiments, and we are able to not only detect and measure \name\ with average AUC score of 0.78-0.96, but also to provide significant insights about the behavior of \name\ entities and the impact on users' privacy.
In the hands of privacy researchers, policy makers and end-users, \fw\ can be an invaluable tool for raising awareness and increasing transparency on tracking practices used by the ad-ecosystem.


\section{Introduction}\label{sec:introduction}

\begin{figure}[t]
	\centering
	\includegraphics[width=0.70\columnwidth]{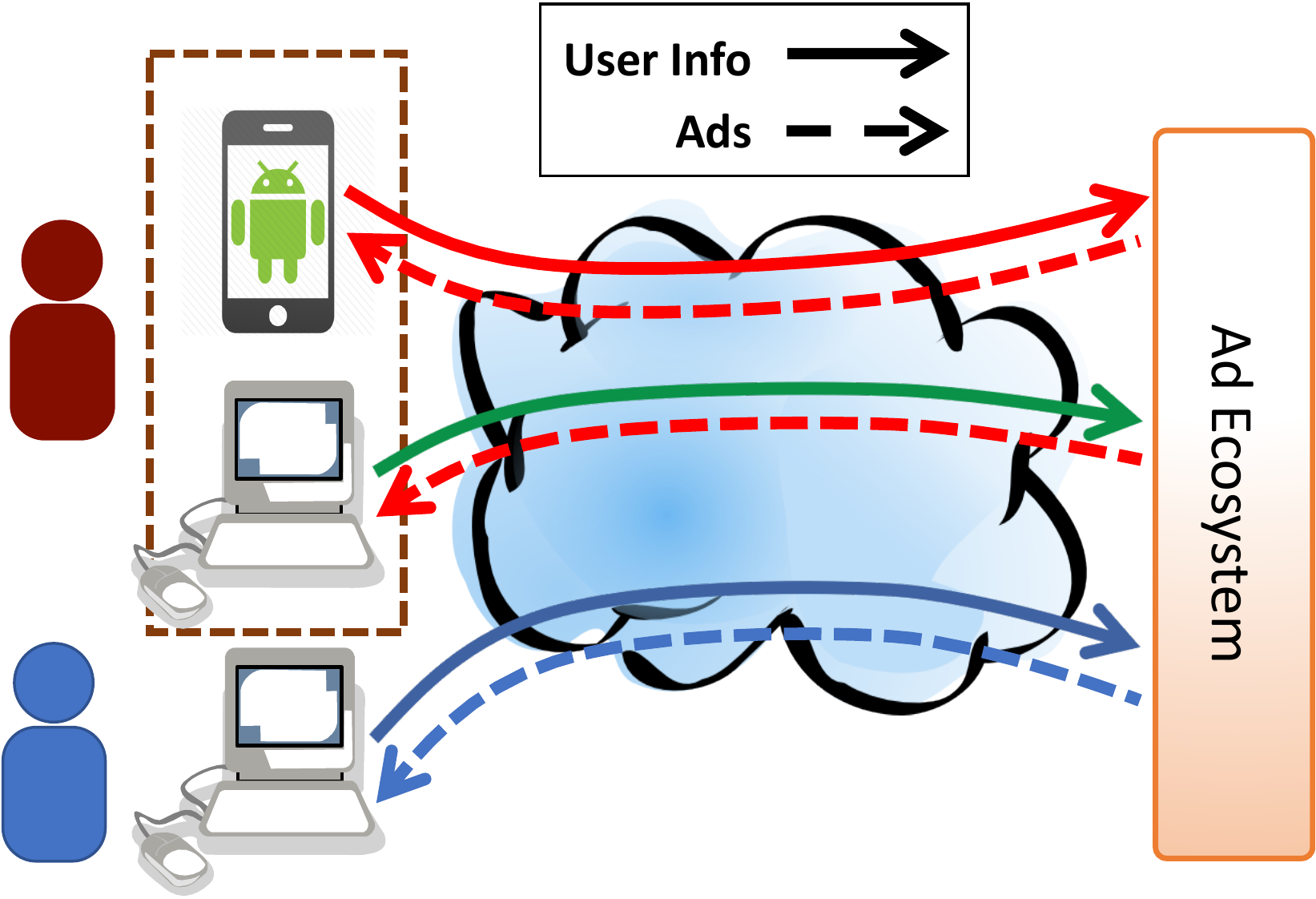}
	\caption{High level representation of \lname.}
	\label{fig:overview}
\vspace{-4mm}
\end{figure}

Online advertising has become a driving force of the economy, with digital ad spending already surpassing the spending for TV-based advertising in 2017~\cite{digital_ad_vs_tv}, and expected to reach \$327 billion in 2019~\cite{digital_ads}.
This is because online advertising can be easily tailored to, and target specific audiences. 
In order to personalize ads, advertisers employ various tracking practices to collect user behavioral and browsing data.



Until recently, the tracking of a user was confined to the physical boundary of each one of her devices.
However, as users typically own multiple devices~\cite{mobile_facts, conversion-attribution}, advertisers have started employing advanced targeting practices specifically designed to track and target users across {\bf all} their devices.
These efforts indicate a radical shift of the ad-targeting paradigm, from \textit{device-centric} to \textit{user-centric}.
In this new paradigm, an advertiser tries to identify which devices (e.g., smartphone, tablet, laptop) belong to the same user, and then target her across all devices with ads related to her overall online behavior.
Figure~\ref{fig:overview} illustrates a typical \lname (\name) scenario, where a user is targeted with relevant ads in her second device (desktop), due to the behavior exhibited to the ad-ecosystem from her first device (mobile).

A recent FTC Staff Report~\cite{ftc2017report} states that \name\ can be deterministic or probabilistic, and companies engaging in such practices typically use a mixture of both techniques.
Deterministic tracking utilizes 1st-party login services that require user authentication (e.g., Facebook, Twitter, Gmail). 
These 1st-party services often share information (e.g., a unique identifier) with 3rd-parties, enabling them to perform a more effective \name.
In the case of probabilistic \name, there are no shared identifiers between the users' devices, and 3rd-parties attempt to identify which devices belong to the same user by considering network access data, common behavioral patterns in browsing history, etc.
In fact, to understand the degree to which \name\ trackers 
appear on the Web, we measured their frequency of appearance on Alexa Top-10k websites: companies performing probabilistic \name\ can be found in $\sim27\%$ of the websites, and when also considering deterministic \name, this coverage reaches $\sim80\%$.
%
Also, several advertising companies such as Criteo~\cite{criteo}, Tapad~\cite{tapad}, Drawbridge~\cite{draw_cdt} etc., 
claim that they can track users across devices 
with very high accuracy (e.g., Drawbridge's Cross-Device connected consumer graph is 97.3\% accurate~\cite{draw_acc}).


%
In spite of its big impact on user privacy, apart from some empirical evidence about \name, there is only a limited work investigating it.
In the most close work to ours, Zimmeck et al.~\cite{zimmeck2017}, designed an algorithm that correlates mobile and desktop devices into pairs by considering devices' browsing history and IP addresses.
While this approach shows that correlation of devices is possible when such data are available, it does not provide an approach for detecting and measuring \name.
In fact, to the best of our knowledge, there is no existing approach to audit the probabilistic \name\ ecosystem and the factors that impact its performance on the Web.
%
%
\emph{
Our work is the first to propose a novel methodology that enables auditing the \name\ ecosystem in an automated and systematic way.
In effect, our work takes the first and crucial step in understanding the inner workings of the \name\ mechanics and measure different parameters that affect how it performs.}

The methodology proposed in this work is based on the following idea: we want to detect when \name\ trackers 
successfully correlate a user's devices, by identifying cross-device targeted behavioral ads they send, i.e., ads that are delivered on one device, but have been triggered because of the user's browsing on a different device.
In order to design this methodology, we first study browsing data of real users with multiple devices from~\cite{zimmeck2017} and extract topics of interest and other user behavioral patterns.
Then, to make trackers correlate the different devices of the end-user and serve cross-device targeted ads, we employ artificially created personas with specific interests, to emulate realistic browsing activity across the user devices as extracted from the real data.

We build \fw, a novel framework that materializes our methodology in order to collect, categorize and analyze all the ads delivered to the different user devices, and evaluate with simple and advanced statistical methods the potential existence of \name.
Through a variety of experiments we are able to measure \name\ with an average AUC of 0.78-0.96.
Specifically, in the simplest experiment, where the user exhibits significant browsing activity mainly from the mobile device, the average value of AUC is 0.78 for the 10 different behavioral profiles used.
When the user exhibits significant browsing activity from both devices (mobile and desktop), with a matching behavioral profile, we observe \name\ with an average AUC of 0.83.
In the case of visiting specifically chosen websites that employ multiple known \name\ trackers, we achieve AUC score of 0.96.
We also find that browsing in incognito can reduce the effect of \name, but does not eliminate it, as trackers can perform device matching based only on the current browsing session of the user, and not all her browsing history.
Finally, we compare the data collected with our real user-driven artificial personas (such as \name\ trackers found, types of ads detected, etc.) with corresponding distributions observed in the real user data from~\cite{zimmeck2017}, offering a strong validation to the realistic design of \fw.

Overall, our main contributions in this work are:
\begin{squishlist}

\item Design a novel, real data-driven methodology for detecting \name\ by triggering behavioral cross-device targeted ads on one user device, according to specifically-crafted emulated personas, and then detecting those ads when delivered on a different device of the same user.
\item Implement \fw, a practical framework for \name\ measurements.
\fw\ has been designed to provide scalability for fast deployment of multiple parallel device instances, to support various experimental setups, and to be easily extensible.
\item Conduct a set of experiments for measuring the potential existence of \name\ in different types of emulated users, with an average AUC score of 0.78-0.96, and investigate the various factors that affect its performance under different classes of experimental setups and configurations.
\end{squishlist}


	\section{Background \& Related Work}
\label{sec:related_work}

In this section, we provide the necessary terminology to understand the technical contributions of our work, and in parallel we present various mechanisms and technologies proposed in related works.

\subsection{Personalized Targeted Advertising}

As the purpose of advertising is to increase
market share,
the advertising industry continuously develops new mechanisms to deliver more effective 
ads. 
These mechanisms involve the delivery of contextual ads, targeted behavioral ads, and also retargeted ads.

Contextual advertising refers to the delivery of ads relevant to the content of the publishing website. 
With regards to the effectiveness of contextual advertisement, Chun et al.~\cite{chun2014} found that it enhances brand recognition and that users tend to have favourable attitudes towards it.
In one of the first works in this area, Broder et al.~\cite{broder_2007} proposed an approach for classifying ads and web pages into a broad taxonomy of topics, and then matching web pages with semantically relevant ads. 
A large body of work also investigates targeted behavioral advertising with regards to different levels of personalization, based on the type of information that is used to target the user~\cite{bleier2015a, aguirre2015, tucker2014,papadopoulos2019cookie,pachilakis19}, and its effectiveness~\cite{ yan_2009, farahat_2012, lewis_2011, gironda2018}. 
Interestingly, Aguirre et al.~\cite{aguirre2015} found that, while highly personalized ads are more relevant to users, they increase users' sense of vulnerability. In another study, Dolin et al. ~\cite{dolin2018unpacking} measured users' comfort regarding personalized advertisement.
In a different direction of investigation, Carrascosa et al.~\cite{carrascosa2015} developed a methodology that employs artificially-created behavioral profiles (i.e., personas) for detecting behavioral targeted advertising at scale. 
Their methodology could distinguish interest-based targeting from other forms of advertising such as retargeting. 
An extensive review of the literature about behavioral advertising can be found in~\cite{boerman2017}.



\subsection{Leakage of Personal Information}

In order to serve highly targeted ads, advertisers employ various, often questionable and privacy intrusive, techniques for collecting and inferring users' personal information. 
They typically employ techniques 
for tracking users visits across different websites, which allow them to reconstruct parts of the users' browsing history. 
Numerous works investigate the various approaches employed by trackers, and focus on protecting users' privacy.

In a recent work, Papadopoulos et al.~\cite{panpap_imc} developed a methodology that enables users to estimate the actual price advertisers pay for serving them ads. 
The range of these prices can indicate which personal information of the user is exposed to the advertiser and the sensitivity of this information. 
Liu et al.~\cite{adreveal} proposed \textit{AdReveal}, a tool for characterizing ads, and found that advertisers frequently target users based on their interests and browsing behavior.
Lecuyer et al.~\cite{lecuyer2014} proposed \textit{XRay}, a data tracking system that allows users to identify which data is being used for targeting, by comparing
outputs from different accounts. 
In another work, they propose \textit{Sunlight}~\cite{lecuyer2015}, a system that employs methodologies from statistics and machine learning to detect targeting at large scale. 

Bashir et al.~\cite{bashir2016} developed a methodology that detects information flows between ad-exchanges. 
This approach leverages retargeted ads, in order to detect when ad-exchanges share the user's information between them, for tracking and retargeting the user.
Datta et al.~\cite{datta2015} developed \textit{AdFisher}, a tool that explores causal connections between users' browsing activities, their ad settings and the ads they receive, and found cases of discriminatory ads. 
This tool uses machine learning to determine, based on the ads received, if the user belongs to a group of users that exhibit a specific browsing behavior i.e., visited specific websites that affected their behavioral profile. 
Castelluccia et al.~\cite{betrayed} showed that targeted ads contain information that enable reconstruction of users' behavioral profiles, and that user's personal information can be revealed to any party that has access the ads received by the user.

In order to enable ad targeting without compromising user privacy, Toubiana et al.~\cite{adnostic} and Guha et al.~\cite{guha2011} proposed \textit{Adnostic} and \textit{Privad}, respectively. 
These two approaches try to protect users' privacy by keeping user profiles on the client-side and thus, hiding user activities and interests from the ad-network.
Furthermore, in an attempt to provide a better alternative, 
Parra-Arnau et al.~\cite{parraarnau2017}, proposes a tool that allows users to control which information can be used for the purpose of advertising.

Furthermore, many works investigate privacy leakage, specifically, in mobile devices and the different factors influencing mobile advertising~\cite{php, terkki2017, razaghpanah, mobilead, meng2016}.
A recent study by Papadopoulos et al.~\cite{php} compared privacy leakage when visiting mobile websites and using mobile apps.
Meng et al.~\cite{meng2016} studied the accuracy of personalized ads served by mobile applications based on the information collected by the ad-networks. 
Also, Razaghpanah et al.~\cite{razaghpanah} developed a technique that detects third-party advertising and tracking services in the mobile ecosystem and uncovers unknown relationships between these services.

\subsection{Web Tracking}

As mentioned previously, various techniques are employed for tracking and correlating users' activities across different websites.
Many works investigated stateful tracking techniques~\cite{roesner2012, olejnik2014, englehardt_ccs, trackers, lerner2016},
and also stateless techniques such as browser fingerprinting~\cite{eckersley2010, acar2013, acar2014, nikiforakis2013, nikiforakis_www15, panchenko2016}. 
One of the first studies about tracking~\cite{mayer2012},
investigated which information is collected by third parties and how users can be identified. 
Roesner et al.~\cite{roesner2012} measured the prevalence of trackers and different tracking behaviors in the web. 

Olejnik et al.~\cite{olejnik2014} investigated ``cookie syncing'', a technique that enables third parties to have a more completed view on the users' browsing history by synchronizing their cookies. 
Acar et al.~\cite{acar2014} investigated the prevalence of ``evercookies'' and the effects of cookie respawning in combination with cookie syncing. 
Englehardt and Narayanan~\cite{englehardt_ccs} conducted a large scale measurement study to quantify stateful and stateless tracking in the web, and cookie syncing,
while Lerner et al.~\cite{lerner2016} conducted a longitudinal measurement study of third party tracking behaviors
and found that tracking has increased in prevalence and complexity over time. 

With regards to stateless tracking, Nikiforakis et al.~\cite{nikiforakis2013} investigated various fingerprinting techniques employed by popular trackers and measured the adoption of fingerprinting in the web. 
Acar et al.~\cite{acar2013} proposed 
\textit{FPDetective}, 
a framework to detect fingerprinting by identifying and analyzing specific events such as the loading of fonts, or accessing specific browser properties. 
In another work, Nikiforakis et al.~\cite{nikiforakis_www15} proposed \textit{PriVaricator}, a tool that employs randomization to make fingerprints non-deterministic, in order to make it harder for trackers 
to link user fingerprints across websites.  
Also, in a recent work, Cao et al.~\cite{cao2017} proposed a 
fingerprinting technique that utilizes OS and hardware level features, 
for enabling user tracking not only within a single browser, but also 
across different browsers on the same machine.

\subsection{Cross-Device Tracking}

A few recent works investigate \lname that is implemented based on technologies such as ultrasound and Bluetooth, and measure the prevalence of these approaches~\cite{mavroudis2017, arp2017, korolova_2018}.
As in this work we focus on web based \lname, our work is complementary to works that investigate such technologies. 


A work by Brookman et al.~\cite{brookman2017}, one of the few that investigate \name\ on the web, provides some initial insights about the prevalence of trackers.
This work examines 100 popular websites in order to determine which of them disclose data to trackers, identifies which websites contain trackers known to employ \name\ techniques, and also investigates if users are aware of
these techniques. 

During the Drawbridge Cross-Device Connection competition of the ICDM 2015 conference~\cite{drb}, the participants were provided with a dataset~\cite{draw_data} that contained information about some users' devices, cookies, IP addresses and also browsing activity, and were challenged to match cookies with devices and users.
This resulted in a number of short papers~\cite{drb_anand, drb_cao, drb_kejela, drb_kim, drb_landry, diazmorales2015, drb_selsaas, drb_wal} that describe different machine learning approaches followed during the competition for matching devices and cookies.
Some of the proposed methods achieved accuracy greater than 90\%, and seen from a different point compared to our work, showed that users' devices can be potentially correlated if enough network and device information is available.

Zimmeck et al.~\cite{zimmeck2017} conducted an initial small-scale exploratory study on \name\ based on the observation of cross-device targeted ads in two ``paired'' devices (mobile and desktop) over the course of two months.
Following this exploration, they collected the browsing history of 126 users, from which 107 have provided data from both their desktop and mobile device, and designed an algorithm that estimates similarities and correlates the devices into pairs. 
This approach, which is based on IP addresses and browsing history, and achieves high matching rates, shows that users' network information and browsing history can be used for correlating user devices, and thus potentially for \name.

In general, research around \name\ is still very limited; in fact, only~\cite{zimmeck2017, brookman2017} initially studied some of its aspects, but without proving its actual existence or providing a methodology for detecting and measuring it. 
Overall, our work builds on these early studies on \name, as well as past studies on detection of web tracking during targeted ads.
We propose the first of its kind methodology for systematic investigation of probabilistic \name, by leveraging artificially-created profiles with specific web behaviors, and measuring the existence of, and factors affecting \name\ in various experimental setups.


\section{A methodology to measure CDT}
\label{sec:overview}

\begin{figure*}[t]
	\centering
	\includegraphics[width=1.70\columnwidth]{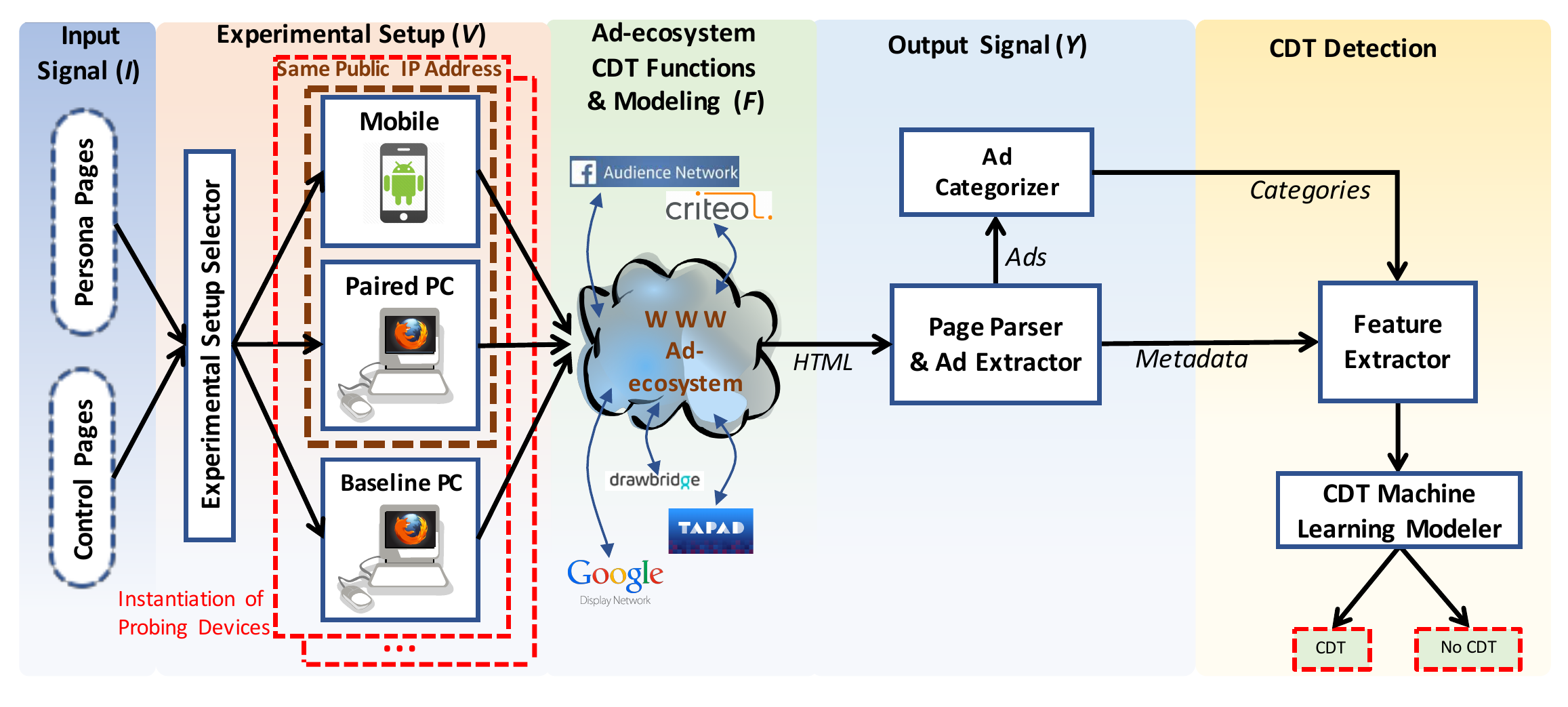}
	\caption{High level representation of methodology design principles and units for \name\ measurements.}
	\label{fig:generic_system_design}
\end{figure*}

The proposed methodology emulates realistic browsing activity of end-users across different devices, and collects 
and categorizes all ads delivered to these devices based on the intensity of the targeting.
Finally, it compares these ads with baseline browsing activity to establish if \name\ is present or not, at what level, and for which types of user interests.

%

\subsection{Design Principle}

In general, the \name\ performed by the ad-ecosystem is a very complex process, with multiple parties involved, and a non-trivial task to dissect and understand.
To infer its internal mechanics, we rely on probing the ecosystem with consistent and repeatable inputs ($\texttt{I}$), under specific experimental settings ($\texttt{V}$), allowing the ecosystem to process and use this input via transformations and modeling ($\texttt{F}$), and produce outputs we can measure on the receiving end ($\texttt{Y}$):
\setlength{\belowdisplayskip}{0pt} \setlength{\belowdisplayshortskip}{0pt}
\setlength{\abovedisplayskip}{0pt} \setlength{\abovedisplayshortskip}{0pt}
\[
\texttt{(I,V)}\xrightarrow[\text{}]{\texttt{F}} \texttt{Y}
\]
In this expression, the unknown $\texttt{F}$ is the probabilistic modeling performed by \name\ entities, allowing them to track users across their devices. 
Following this design principle, our methodology allows to push realistic input signals to the ad-ecosystem via website visits, and measure the ecosystem's output through the delivered ads, to demonstrate if \texttt{F} enabled the ecosystem to perform probabilistic \name. 
An overview of our methodology is illustrated in Figure~\ref{fig:generic_system_design}.

\subsection{Design Overview}\label{sec:meth_overview}
\subsubsection{Input Signal \textit{(I)}}

To trigger \name, we first need to inject to the ad-ecosystem some activity from a user's browsing behavior ($\texttt{I}$).
This \textit{input} can be visits (i) to pages of interest (e.g., travel, shopping), or (ii) to control pages of null interest (e.g., weather pages).
Intuitively, the former can be used first to demonstrate particular behavior of a user from a given device (mobile), and the latter afterwards for collecting ads delivered as the \textit{output} of the ecosystem ($\texttt{Y}$) due to $\texttt{I}$, to that device, or other device of the same user (desktop).

\noindent
\textbf{Persona Pages.}
We extract real users' interests from the dataset provided by Zimmeck et al.~\cite{zimmeck2017} and leverage an approach similar to Carrascosa et al.~\cite{carrascosa2015} to emulate browsing behavior according to specific web categories, and create multiple, carefully-crafted \textit{personas} of different granularities. 
This design makes the methodology systematic and repeatable and produces realistic browsing traffic from scripted browsers.
For each persona, our approach identifies a set of websites (dubbed as \textit{persona pages}) that have, at the given time, active ad-campaigns.
This ``\textit{training activity}'' aims to drive \name\ trackers into possible \textit{device-pairing} between the user's two devices with high degree of confidence.

\noindent
\textbf{Control Pages}.
Following past works~\cite{carrascosa2015, bashir2016}, all devices in the system collect ads by visiting neutral websites that typically serve ads not related to their content, thus, reducing bias from possible behavioral ads delivered to specific type of websites.
We refer to these websites as \textit{control pages}.
We detail the design of personas and control pages in~\cref{sec:input}.

\noindent
\subsubsection{Experimental Setup \textit{(V)} }
\textbf{No 1st-party logins.}
Since we focus on probabilistic \name, we assume that the emulated user does not visit or log into any 1st-party service that employs deterministic \name\ and thus, there is no common identifier (e.g., email address, social network ID) shared between the user's devices.

\noindent
\textbf{Devices, IP addresses \& Activity.}
The approach we follow 
is based on triggering and identifying behavioral cross-device targeted ads, and specifically ads that appear on one of the user's devices, but have been triggered by the user's activity on a different device.
For this trigger to be facilitated, the ad-ecosystem must be provided with hints that these two devices belong to the same user.
Zimmeck et al.~\cite{zimmeck2017} suggest that in many cases, the devices' IP address is adequate for matching devices that belong to the same user. Also, according to relevant industrial teams~\cite{lotame_pairing,adelphic_pairing} more signals can be used, such as location, browsing, etc., for device matching.

Following these observations, our methodology requires a minimum of three different devices: one mobile device and two desktop computers, with two different public IP addresses.
We assume that two devices (i.e., the mobile and one desktop) belong to the same user, and are connected to the same network.
That is, these devices have the same public IP address, are active in the same geolocation as in a typical home network, and will be considered by the ad-ecosystem as producing traffic from the same user.
The second desktop (i.e., \textit{baseline PC}), which has a different IP address, is used for receiving a different flow of ads while replicating the browsing of the user's desktop (i.e., \textit{paired PC}).
This control instance is used for establishing a baseline set of ads to compare with the ads received by the user's paired PC.

\noindent
\textbf{CDT Direction.}
In principle, the design allows the investigation of both directions of  \name.
That is, users may first browse on the mobile device, and then move to their desktop, and vice versa.
However, since ad-targeting companies such as AdBrain and Criteo support that the direction from mobile to desktop is more suitable for cross-device retargeting~\cite{adwords_device, AdBrain, criteo_cdt}, 
in this work we focus on the mobile to desktop direction ($Mob \rightarrow PC$).
In essence, the mobile device performs a specifically instructed web browsing session to establish the persona, by visiting the set of \textit{persona pages}, i.e., \textit{training} phase; then, the two desktop computers perform web browsing, i.e., \textit{testing} phase, where they visit the set of \textit{control pages} and collect the delivered ads.
The browsing performed by the desktops is synchronized by means of visiting the same pages and performing the exact same clicks.

\subsubsection{Output Signal \textit{(Y)}}
In order to handle the Output Signal and transform it appropriately, we design and implement two different components: (i) Page Parser \& Ad Extractor and (ii) Ad Categorizer.
The first is responsible for the identification and extraction of ad elements inside the webpages.
The module uses string matching techniques and a public list of common ad-domains (\textit{Easylist}~\cite{easylist}) to identify the delivered ads. 
The second module assigns a keyword on each ad identified on the previous step, based on its type and content (e.g., ``Online Shopping", ``Fashion", ``Recreation", etc.).
Using both modules, we store the ads delivered in all devices of our experimental setup along with their categories, as well as data related to the activity of the devices that attracted these ads.

\subsubsection{\name\ Detection}
\textbf{Comparing Signals.} 
Various statistical methods can be used to associate the input signal $\texttt{I}$ of persona browsing in the mobile device, with the output signal $\texttt{Y}$ of ads delivered to the potentially paired-PC.
For example, simple methods that perform similarity computation between the two signals in a given dimensionality (e.g., Jaccard, Cosine) can be applied.
These methods, as well as typical statistical techniques (e.g., permutation tests) capture only one dimension of each input/output signal and thus, might not be suitable for measuring with confidence the high complexity of the \name\ signal.
In this case, more advanced methods can be employed, such as Machine Learning techniques (ML) for classification of the signals as similar enough to match, or not. 
In our analysis, we mainly focus on ML to compute the likelihood of the two signals being the product of \name, as it takes into consideration this multidimensionality in the feature space.
We describe the modeling and methods used for ML in~\cref{sec:cdtmodel}.


\section{Framework  Implementation}
\label{sec:methodology}

A high level overview of our methodology, and its materialization by our framework \fw, is presented in Figure~\ref{fig:generic_system_design} and described in~\cref{sec:overview}.
In the following, we provide more details about its building blocks, and argue for various design decisions taken while implementing this methodology into the fully-fledged automated system.

\subsection{Input Signal: Control Pages  \& Personas}\label{sec:input}

\begin{figure}[t]
	\centering
	\includegraphics[width=0.9\columnwidth]{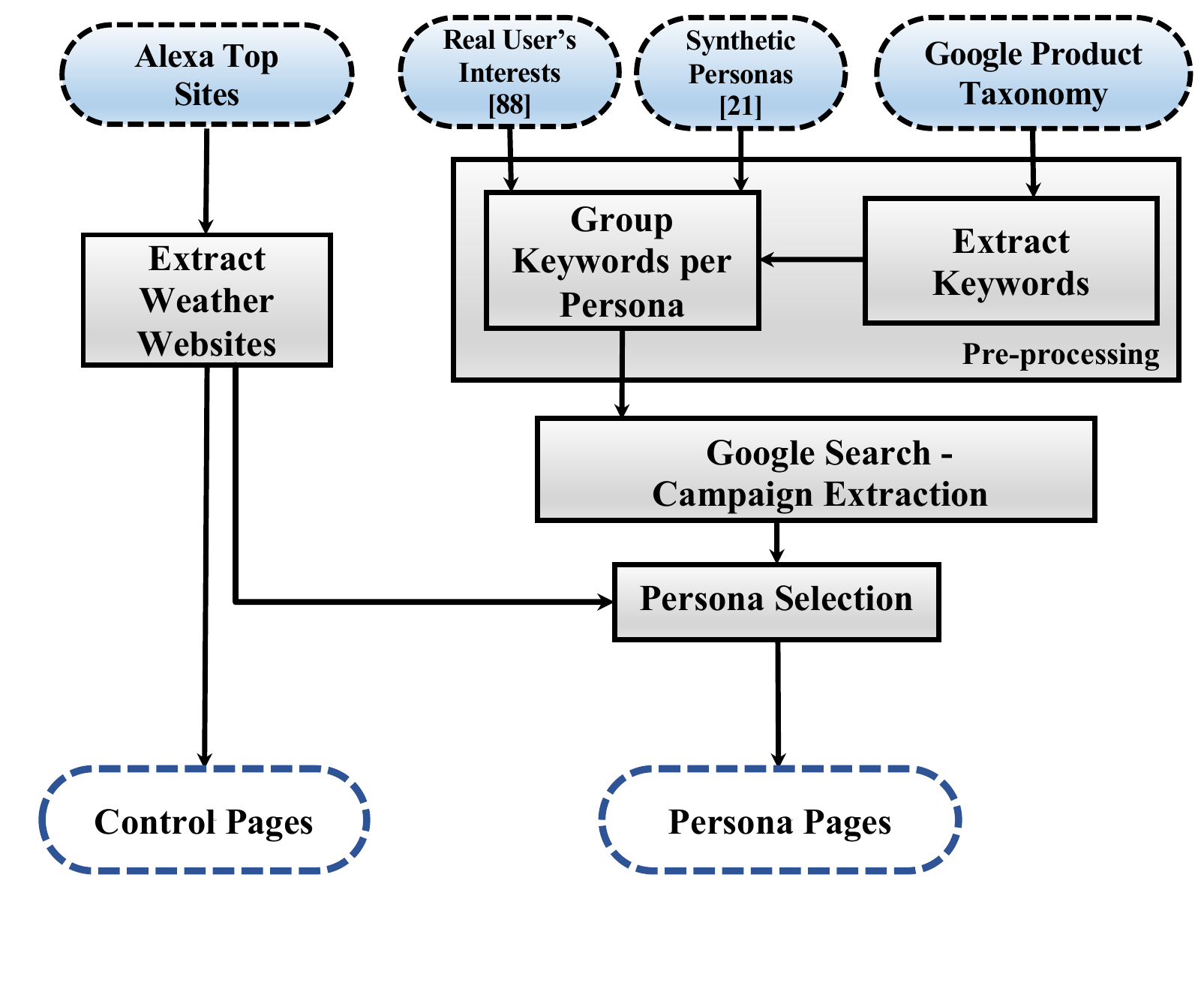}
	\caption{Persona design and automatic generation.}
	\label{fig:personas}
\end{figure}

\textbf{Persona Pages.} A critical part of our methodology is the design and automatic building of realistic user personas.
Each persona has a unique collection of visiting links, that form the set of \textit{persona pages}.
Since we do not know in advance which e-commerce sites are conducting cross-device ad-campaigns, we design a process to dynamically detect active persona pages of given interest categories. 
Our approach for persona generation is shown in Figure~\ref{fig:personas}.

We first use the list of topics of Zimmeck at al.~\cite{zimmeck2017}, that describe real user's online interests. 
We perform a clustering based on the content of each interest and label the clusters appropriately (e.g., we group together: ``Shopping" and ``Beauty and Fashion" under the label: ``Shopping and Fashion").
Then, we use the persona categorization of Carrascosa et al.~\cite{carrascosa2015} for their top 50 personas, and select only those personas that describe similar interests with the previously formed list.
For the resulting intersection of personas from the two lists, we iterate through the Google Product Taxonomy list~\cite{g_taxonomy} to obtain the related keywords for each one.

For increasing the probability to capture active ad-campaigns that can potentially deliver ads to the devices, we use Google Search as it reveals campaigns associated with products currently being advertised.
That is, if a user searches for specific keywords (e.g., ``men watches''), Google will display a set of results, including sponsored links for sites conducting campaigns for the terms searched.
In this way, we use the keywords set for each persona, as extracted above, and transform them into search queries by appending common string patterns such as ``buy'', ``sell'', and ``offers''. This process is repeated until between five and ten unique domains per persona are collected. 
If the procedure fails, no persona is formed.

As the effectiveness of a persona depends on the active ad-campaigns at the given time, 
in our experiments, we deploy personas in 10 categories related to shopping, traveling, etc. (full list shown in Table~\ref{table:personas}).
With this procedure, we manage to design personas similar enough with real users, as well as with emulated users designed in previous works~\cite{carrascosa2015,bashir2016,bashir2019quantity,zimmeck2017}.
\begin{table}[t!]
	\centering
	\caption{Behavioral personas used in our experiments.}
	\label{table:personas}
	\small
	\begin{tabular}{ c|l } 
		\textbf{Persona} & \textbf{Category - Description} \\
		\toprule
		\rowcolor{Gray}
		1 & Online Shopping - Accessories, Jewelry.\\
		2 & Online Shopping - Fashion, Beauty.\\
		\rowcolor{Gray}
		3 & Online Shopping - Sports and Accessories.\\
		4 & Online Shopping - Health and Fitness.\\
		\rowcolor{Gray}
		5 & Online Shopping - Pet Supplies.\\
		6 & Air Travel.\\
		\rowcolor{Gray}
		7 & Online Courses and Language Resources.\\
		8 & Online Business, Marketing , Merchandising.\\
		\rowcolor{Gray}
		9 & Browser Games - Online Games.\\
		10 & Hotels and Vacations.\\
		\bottomrule
	\end{tabular}
\end{table}

\noindent
\textbf{Control Pages.} For retrieving the delivered ads (after any type of browsing), we employ a set of webpages that contain: (i) easily identifiable ad-elements and (ii) a sufficient number of ads that remains consistent through time.
These pages have neutral context and do not affect the behavioral profile of the device visiting them.
For most of the experiments in \cref{sec:evaluation}, we use a set of five popular weather websites\footnote{\url{accuweather.com}, \url{wunderground.com}, \url{weather.com}, \url{weather-forecast.com}, \url{metcheck.com}} 
as control pages, similarly to~\cite{carrascosa2015}.
We manually confirmed the neutrality of these pages, by observing no contextual ads delivered to them.
When visiting the set of control pages, our methods extract and categorize all the ads received, in order to identify those that have been potentially resulted from \name.

\subsection{Experimental System Setup}\label{sec:systemset}

The experimental setup contains different types of units, connected together for replicating browsing activity on multiple devices.
Typically, \name\ is applied on two or more devices that belong to the same user, such as a desktop and a mobile device.
Thus, the system contains emulated instances of both types, controlled by a number of experimental parameters.\\
\noindent
\textbf{Devices \& Automation.}
The desktop devices are built on top of the web measurement framework OpenWPM~\cite{englehardt_ccs}.
This platform enables launching instances of the Firefox browser, performs realistic browsing with scrolling, sleeps and clicks, and collects a wide range of measurements in every browsing session.
It is also capable of storing the browser's data (cookies, local cache, temporary files) and exports a browser profile after the end of a browsing session, which can be loaded in a future session.
With these options, we can perform \textit{stateful} experiments, as a typical user's web browser that stores all the data through time, or \textit{stateless} experiments to emulate browsing in incognito mode. 

For the mobile device, we use the official Android Emulator~\cite{avd},
as well as the Appium UI Automator~\cite{appium} for the automation of browsing.
We build the mobile browsing module on top of these components to automate visits to pages via the Browser Application.
This browsing module provides functionalities for realistic interaction with a website, e.g., scrolling, click and sleep rate.
Similarly to the desktop, it can run either in a \textit{stateful} or \textit{stateless} mode.

\noindent
\textbf{Experimental Setup Selector.}
As shortly described in~\cref{sec:overview}, we need two phases of browsing to different types of webpages (\textit{training} and \textit{testing}), in order to successfully measure \name.
For that reason, we set the two browsing phases in the following way:
During the training phase, the selected device visits the set of \textit{Persona Pages} for a specific duration, referred to as training time (\textit{$t_{train}$}).
The test phase is the set of visits to \textit{control pages} for the purpose of collecting ads.
During this phase, we control the duration of browsing (i.e., \textit{$t_{test}$}).
The experimental setup selector controls various parameters such as: which type of device will be trained and tested, the times $t_{train}$ and $t_{test}$, the sequence of time slots for training and testing from the selected device, number of repetitions of this procedure, etc.

\begin{figure}[t]
	\centering
	\includegraphics[width=0.80\columnwidth]{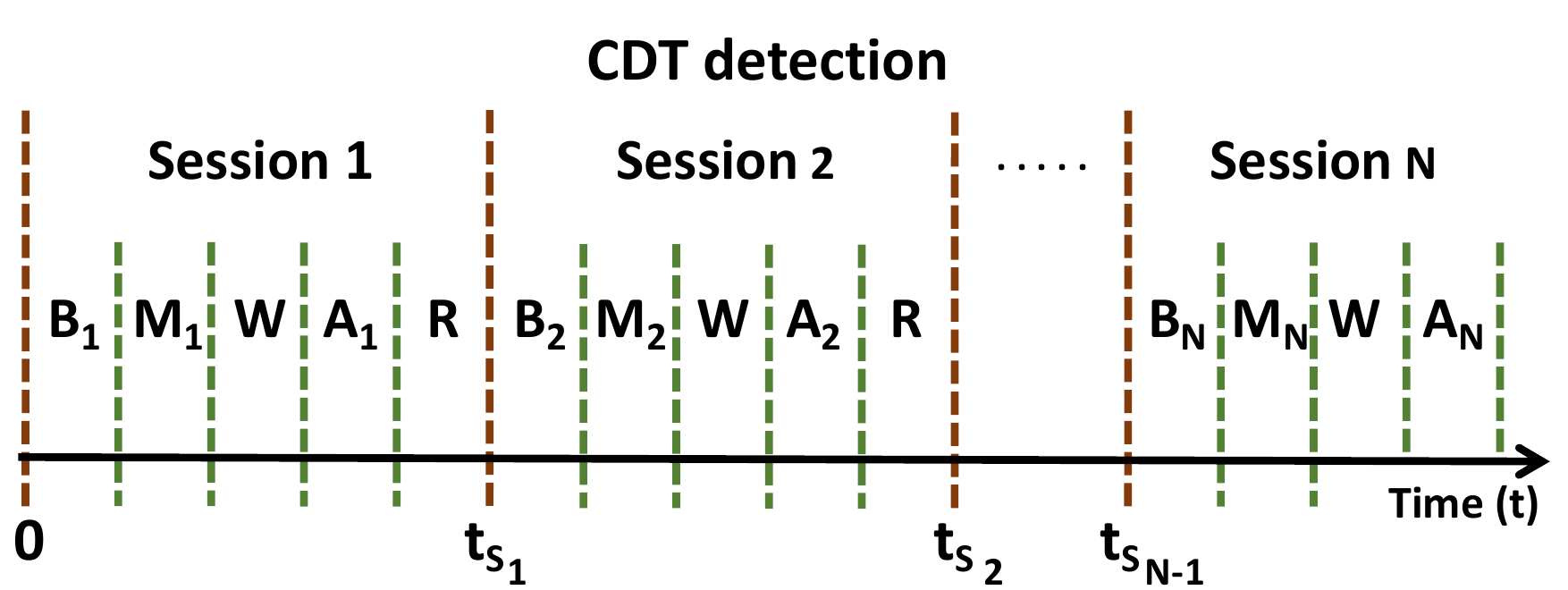}
	\caption{Timeline of phases for CDT measurement.\\
		$M_i$: mobile training time $t_{train}$ + testing time $t_{test}$;\\
		$B_i$($A_i$): desktop testing time $t_{test}$ before (after) mobile phase;\\
		W: wait time ($t_{wait}$); R: rest time ($t_{rest}$); $t_{S_i}$: time of session $i$.}
	\label{fig:cdt_detection}
\end{figure}

\noindent
\textbf{Timeline of phases.}
Each class of experiments is executed multiple times (or runs), through parallel instantiations of the user devices within the framework (as shown in Figure~\ref{fig:generic_system_design}).
Each experimental run is executed following a timeline of phases as illustrated in Figure~\ref{fig:cdt_detection}.
This timeline contains \textit{N} sessions with three primary stages in each: Before, Mobile, and After.
The \textit{Before ($B_i$)} stage is when the two desktop devices perform a parallel test browsing, with a duration of $t_{test}$ time, to establish the state of ads before the mobile device injects signal into the ad-ecosystem.
The \textit{Mobile ($M_i$)} stage is when the mobile device performs a training browsing for $t_{train}$ time, and a test browsing for $t_{test}$ time.
This phase injects the signal from the mobile during training with a persona, but also performs a subsequent test with control pages to establish the state of ads after the training.
Finally, the \textit{After ($A_i$)} stage is when the two desktops perform the final test browsing, with the same duration $t_{test}$ as in \textit{Before ($B_i$)} stage, to establish the state of ads after the mobile training.

After extensive experimentation, we found that a minimum training time $t_{train}$=15 minutes and testing time $t_{test}$=20 minutes are sufficient for injecting a clear signal over noise, from the trained device to the ad-ecosystem.
There is also a waiting time ($t_{wait}$=10 minutes) and resting time ($t_{rest}$=5 minutes) between the stages of each session, to allow alignment of instantiations of devices running in parallel during each session.
In total, each session lasts 1.5 hours and is repeated $N$=15 times during a run.
Through the experimental setup selector, we define the values of such variables ($t_{train}$, $t_{test}$, $t_{wait}$, $t_{rest}$, $N$, type of device), offering the researcher the flexibility to experiment in different cases of \name.

\subsection{Output Signal}\label{sec:pageparser}

\noindent
\textbf{Page Parser.}
This component is activated when the visited page is fully loaded and no further changes occur on the content. 
To collect the display ads, we first need to identify specific DOM elements inside the visited webpages.
This task is challenging due to the dynamic Javascript execution and the complex DOM structures generated in most webpages.
For the reliable extraction of ad-elements and identification of the landing pages,\footnote{Destination websites the user is redirected to when clicking on the ads.} we follow a methodology similar to the one proposed in~\cite{adreveal}.
The functionality of this component is to parse the rendered webpage and extract the attributes of display ads, which also contain the landing pages.

\noindent
\textbf{Ad Extractor.}
In most modern websites, the displayed ads are embedded in \textit{iFrame} tags that create deep nesting layers, containing numerous and different types of elements.
However, the ads served by the control pages are found directly inside the iFrames so the module does not have to handle such complex behavior.
Therefore, the module firstly identifies all the active iFrame elements and filters out the invalid ones that have either empty content or zero dimensions.
Then, it retrieves the \textit{href} attributes of image and flash ads and parses the URLs, while searching for specific string patterns such as \textit{adurl=, redirect=}, etc.
These patterns are typically used by the ad-networks for encoding URLs in webpages.
Next, the module forms the list of candidate landing pages, which are then processed and analyzed to create the set of true landing pages.
The Ad Extractor is fully compatible with the crawlers, and does not need to perform any clicks on the ad-elements, since it extracts only the landing pages' URLs directly from the rendered webpage.
After collecting the candidate landing pages, the module filters them with the \textit{EasyList}~\cite{easylist}, similarly to previous works~\cite{bashir2016,englehardt_ccs}, and stores only the true active ad-domains.
Finally, the Page Parser \& Ad Extractor module also stores \textit{metadata} from the crawls such as: time and date of execution, number of identified ads, number of categories, type and phase of crawl, etc.

\noindent
\textbf{Ad Categorizer.}
To associate landing pages or browsing URLs with web categories, we employ the McAfee TrustedSources database~\cite{mcafee}, which provides URLs organized into categories.	
This system was able to categorize 96\% of the landing pages of our collection into a total of 76 unique categories, by providing up to four semantic categories for each page, while the remaining  4\%  domains were manually classified to the categories above.
The final output contains the landing pages of collected ads, along with their categories.

\subsection{CDT Detection}\label{sec:cdtmodel}

\begin{table}[t!]
	\centering
	\caption{Description of features used by datasets. The type of desktop crawl  values are in  range \{0,1\}, where 0 represents the before/test sessions, while 1   the after/train sessions. The time of crawl is divided in 30 minutes timeslots and is encoded in range \{0,48\}. The day of crawl is encoded in range \{1,7\}. \textit{V}  represents the (enumerated) vectors of values in the sets of: landing pages, training pages, ads and ad categories.}
	\label{appendix:features}
	\resizebox{0.95\columnwidth}{!}{%
	\begin{tabular}{ c|l } 
		\textbf{Feature Label} & \textbf{Description} \\
		\toprule
		\rowcolor{Gray}
		Crawl\_Type &The type of desktop crawl.\\
		Run\_ID & The indexed number of run\{1,4\}.\\
		\rowcolor{Gray}
		Session\_ID& The index of session\{1,15\}.\\
		
		Persona\_Keywords& V: keyword categories of training pages.\\
	\rowcolor{Gray}

		Mobile\_Timeslot & Time of crawl (Mobile).\\
		 Desktop\_Timeslot & Time of crawl (Desktop).\\
		 	\rowcolor{Gray}

		 Desktop\_Day & The day of crawl (Desktop).\\
		\rowcolor{Gray}
		
		Mobile\_Number\_of\_Ads &  \# ad domains (Mobile).\\
		Desktop\_Number\_of\_Ads&  \#  ad domains collected (Desktop).\\
				\rowcolor{Gray}

		Mobile\_Unique\_Number\_of\_Ads & \# distinct ad domains (Mobile).\\
		Desktop\_Unique\_Number\_of\_Ads &\# ad domains (Desktop).\\
			\rowcolor{Gray}
		Mobile\_Number\_of\_Keywords & \# ad categories  (Mobile).\\
	 Desktop\_Number\_of\_Keywords & \# ad categories  (Desktop).\\
				\rowcolor{Gray}

		Mobile\_Unique\_Number\_of\_Keywords & \#  distinct ad categories (Mobile).\\
		Desktop\_Unique\_Number\_of\_Keywords& \# distinct ad categories (Desktop).\\
		\rowcolor{Gray}
		Mobile\_Keywords & V: keyword categories of landing pages (Mobile).\\
		Desktop\_Keywords& V: keyword categories  of landing pages (Desktop).\\
				\rowcolor{Gray}

		Mobile\_Landing\_Pages& V: landing pages of delivered ads (Mobile).\\ 	
		Desktop\_Landing\_Pages& V: landing pages of delivered ads (Desktop.)\\ 	
		\bottomrule
	\end{tabular}
	}
\end{table}

\noindent
Probabilistic \name\ is a kind of task generally suitable for investigation through ML.
Previous work~\cite{zimmeck2017} and industry directions~\cite{lotame_pairing,adelphic_pairing} claim that probabilistic device-pairing is based on specific, well-defined signals such as: IP address, geolocation, type and frequency of browsing activity. 
Since we control these parameters in our methodology, by definition we construct the ground truth with our experimental setups. 
That is, we control (i)~the devices used, which are potentially paired under a given IP address, geolocation and browsing patterns, (ii)~the control instance of baseline desktop device, and (iii)~the browsing with the personas.

Before applying any statistical method, every instance of the input data has to be transformed into a vector of values; each position in the vector corresponds to a feature.
Features are different properties of the collected data: browsing activity of a user during training time, experimental setup used (persona, etc.), time-related details of the experiment, as well as information about the collected ads, which is the output signal received from the given browsing activity.
These features can be studied systematically to identify statistical association between the input and output signals, given an experimental setup.
In effect, our feature space is comprised of a union of these vectors, since all features are either controlled, or measurable by us (detailed description of the features is given in  Table~\ref{appendix:features}).
The only unknown is whether the ad-ecosystem has successfully associated the devices, and if it has exhibited this in the output signal via ads.\\
\noindent
\textbf{One Dimension Statistical Analysis.}
At the first level of analysis, to measure the similarity of distribution of ads delivered in the different devices, 
we compare the signals using a two-tailed permutation test and reject the null hypothesis that the frequency of ads delivered (for a given category) comes from the same distribution, if the t-test statistic leads to a p-value smaller than a significance level $\alpha < 0.05$.
\noindent
\textbf{Multidimensional Statistical Analysis.}
Given that a uni-dimensional test such as the previous one does not take into account the various other features available in each experiment, we further consider ML, which take into account multidimensional data, to decide if the ads delivered in each device are from the same distribution or not.
We transform the problem of identifying if the previously exported vectors are similar enough, into a typical binary classification problem, where the predicted class describes the existence of pairing or not, that may have occurred between the mobile device and one of the two desktop devices.
As a paired combination we consider the desktop device that exists under the same IP address with the mobile device.
The ``not paired'' combination is the mobile device and the baseline desktop.
The  analysis is based on three classification algorithms with different dependences on the data distributions.
An easily applied classifier that is typically used for performance comparison with other models, is the Gaussian Naive Bayes classifier.
Logistic Regression is a well-behaved classification algorithm that can be trained, as long as the classes are linearly separable.
It is also robust to noise and can avoid overfitting by tuning its regularization/penalty parameters.
Random Forest and Extra-Trees classifiers, construct a multitude of decision trees and output the class that is the mode of the classes of the individual trees. 
Also they use the Gini index metric to compute the importance of features.

A fundamental point when considering the performance evaluation of ML algorithms is the selection of the appropriate metrics.
Pure Accuracy can be used, but it's not representative for our analysis, since we want to report the most accurate estimation for the number of predicted paired devices, while at the same time measure the absolute number of miss-classified samples overall.
For this reason, metrics like Precision, Recall and $F_1$-score, and the Area Under Curve of the Receiver Operating Curve (\textit{AUC})
are typically used, since they can quantify this type of information.

\begin{table}[t!]
	\centering
	\caption{Characteristics of the datasets used in each setup (S) of experiments.
	S=\{1,2,3\} are the setups of experiments in ~\cref{sec:shorttrain}, ~\cref{sec:longtrain} and~\cref{sec:stateless}, respectively;
	$t_{total}$: the total duration of experiment;
	$t_{train}$: the training duration;
	$t_{test}$: the testing duration;
	I: independent personas;
	C: data combined from personas;
	SF: stateful browser;
	SL: stateless browser;
	B: boosted CDT browsing.}
	\label{appendix:datasets}
	\resizebox{1.03\columnwidth}{!}{%
	\begin{tabular}{clcccccc}
		\textbf{S}	&	\textbf{Personas}	&	\textbf{Runs}	&	\textbf{$t_{train}$}	&	\textbf{$t_{test}$}	&	\textbf{$t_{total}$}	&	\textbf{Samples}&	\textbf{Features}	\\
		\toprule
		\rowcolor{Gray}
			1a	&	10 (I, SF)		&	4		&	15min	&	20min	&	37 days&	240		&	1100		\\
		
			1b	&	10 (C, SF)		&	-		&	-		&	-		&	-	&	2400		&	2201	\\

		\rowcolor{Gray}
			2a	&	2 (I, SF)			&	4		&	480min	&	30min	&	6 days	&	192		&	600		\\
				
			2b	&	2 (C, SF)			&	-		&	-		&	-		&	-	&	384		&	750 	\\
				\rowcolor{Gray}
			2c	&	2 (I, SF, B)		&	4		&	480min	&	30min	&	6 days 	&	192		&	500		\\
			2d	&	2 (C, SF, B)		&	-		&	-		&	-		&	-	&	384		&	576		\\
		\rowcolor{Gray}
			3a	&	5 (I, SL)			&	2		&	15min	&	20min	&	9 days	&	120		&	450	\\
			3b	&	5 (C, SL)			&	-		&	-		&	-		&	-	&	600		&	880		\\
		\bottomrule
	\end{tabular}
}

\end{table}


\section{Experimental Evaluation}
\label{sec:evaluation}


We use the \fw\ framework to perform various experiments and construct different datasets for each.
Since every experimental setup has different experimental parameters (i.e., training and testing time,  number of personas,  browsing functionalities), the datasets vary in terms of samples size and feature space.
The datasets collected during our experiments and used in our analysis are presented in the Table~\ref{appendix:datasets}.

\subsection{Does IP-sharing allow \name?}
\label{sec:prel_exp}
A first set of preliminary experiments were performed to demonstrate that our platform can (i)~successfully identify and collect the ads delivered to our multiple devices (mobile and desktops), (ii)~inject browsing signal from a device, thus biasing it to have a realistic persona and (iii)~lead to matching/pairing of devices, which could be due to same behavioral ads, retargeting ads or \name.

First, we use a simple experimental setup: we connect three instances of desktop devices and one mobile device under the same IP address.
We create one persona (as in~\cref{sec:input}), with an interest in ``Online Shopping-Fashion, Beauty'', and following the described timeline of phases, we run this experiment for two days.
Then, we perform one-dimensional statistical analysis, as introduced in~\cref{sec:cdtmodel}, and find that there is no similarity between the mobile with any of desktop devices (null hypothesis rejected with highest p-value=0.030), while all desktop distributions are similar to each other (null hypothesis accepted with lowest p-value=0.33).
These statistical results indicate that there is no clear device-pairing (at the level of ad distribution for the given persona), and that we should consider controlling more factors to instigate it.


Consequently, we expand this experiment by also training one of the desktop devices using the same persona as with mobile.
By repeating the same statistical tests, we find that the mobile and desktop with the same browsing behavior receive ads coming from the same distribution (null hypothesis accepted with lowest p-value=0.84), while the other desktop devices show no similarity with each other or the mobile (null hypothesis rejected with highest p-value=0.008).
This result indicates that browsing behavior under a shared IP address can boost the signal towards advertisers, which they can use to apply advanced targeting, either as \name, or retargeting on each device or a mixture of both techniques.

Finally, these preliminary experiments and statistical tests provide us with evidence regarding the effectiveness of our framework to inject enough browsing signal from different devices under selected personas. 
Our framework is also able to collect ads delivered between devices, that can be later analyzed and linked back to the personas.
Those are fundamental components for our system and importantly they are potentially causing \name\ between the devices involved.
Next, we present more elaborate experimentations with our framework, in order to study \name\ in action.

%

\subsection{Does short-time browsing allow \name?} 
\label{sec:shorttrain}

\noindent
\textbf{Independent Personas: Setup 1a.}
This experimental setup emulates the behavior of a user that browses frequently about some topics, but in short-lived sessions in her devices.
Given that most users do not frequently delete their local browsing state, this setup assumes that the user's browser stores all state, i.e., cookies, cache, browsing history.
This enables trackers to identify users more easily across their devices, as they have historical information about them.
In this setup, every experimental run starts with a clean browser profile; cookies and temporary browser files are stored for the whole duration of the experimental run (stateful).
We use all personas of Table~\ref{table:personas}, and the data collection for each lasts 4 days.

We perform the same statistical analysis as in~\cref{sec:prel_exp}, and find that in 4/10 personas, the mobile and paired desktop ads are similar (null hypothesis accepted with lowest p-value=0.13), while the mobile and baseline desktop ad distributions are different (null hypothesis is rejected with highest p-value=0.009).
This inconsistency is reasonable since the statistical analysis is based only on one dimension (the frequency count of types of ads appearing in the devices), which may not be enough for fully capturing the existence of device-pairing.
For this reason, we choose to use more advanced, multidimensional ML methods which take into account the various variables available, to effectively compare the potential \name\ signals received by the two devices.
\begin{table}[t]
	\centering
	\caption{Performance evaluation for Random Forest in Setups 1a and 1b.
	Left value in each column is the score for Class 0 (C0=\textit{not paired desktop});
	right value for Class 1 (C1=\textit{paired desktop}).}
	\label{tab:persona_eval_shorttrain}
	\small
	\resizebox{1.03\columnwidth}{!}{
	\begin{tabular}{c|cc|cc|cc|c}
		\toprule
		{\textbf{Persona}}&\multicolumn{2}{c|}{\textbf{Precision}}&\multicolumn{2}{c|}{\textbf{Recall}}&\multicolumn{2}{c|}{\textbf{$F_1$-Score}}&\multirow{2}{*}{\textbf{AUC}}\\
		\cline{2-3}\cline{4-5}\cline{6-7}
		(Setup) & \textbf{C0}&\textbf{C1} & \textbf{C0}&\textbf{C1}	& \textbf{C0}&\textbf{C1} &\\		
		\toprule
		\rowcolor{Gray}
		1 (1a) & 0.89 & 0.60 & 0.57 & 0.90 & 0.70 & 0.72 &\textbf{0.73}\\
		2 (1a) & 0.84 & 0.78 & 0.81 & 0.82 & 0.82 & 0.80 &\textbf{0.82}\\
		\rowcolor{Gray}
		3 (1a) & 0.81 & 0.73 & 0.78 & 0.76 & 0.79 & 0.74 &\textbf{0.76}\\
		4 (1a) & 0.87 & 0.78 & 0.87 & 0.78 & 0.87 & 0.78 &\textbf{0.82}\\
		\rowcolor{Gray}
		5 (1a) & 0.94 & 0.65 & 0.68 & 0.93 & 0.79 & 0.76 &\textbf{0.80}\\
		6 (1a) & 0.57 & 0.67 & 0.81 & 0.38 & 0.67 & 0.48 &\textbf{0.59}\\
		\rowcolor{Gray}
		7 (1a) & 0.81 & 0.87 & 0.89 & 0.76 & 0.85 & 0.81 &\textbf{0.81}\\
		8 (1a) & 0.86 & 0.85 & 0.89 & 0.81 & 0.87 & 0.83 &\textbf{0.84}\\
		\rowcolor{Gray}
		9 (1a) &  0.74 & 0.90 & 0.91 & 0.73 &0.82 & 0.81 &\textbf{0.81}\\
		10 (1a) &  0.77 & 0.85 & 0.81 & 0.81 &0.79 & 0.83 &\textbf{0.81}\\ \hline
		combined (1b) &  0.77 & 0.84 & 0.81 & 0.84 & 0.82 & 0.84 &\textbf{0.89}\\		
		\bottomrule
	\end{tabular}
}
\end{table}

The classification results of the Random Forest (best performing) algorithm are reported in Table~\ref{tab:persona_eval_shorttrain}.
We use AUC score as the main metric in our analysis, since the ad-industry seems to prefer higher Precision scores over Recall, as the False Positives have greater impact on the effectiveness of ad-campaigns.\footnote{Tapad~\cite{tapad_cdt} mentions: ``Maintaining a low false positive rate while also having a low false negative rate and scale is optimal. This combination is a strong indicator that the Device Graph in question was neither artificially augmented nor scrubbed.''}
As shown in Table~\ref{tab:persona_eval_shorttrain}, the model achieves high AUC scores for most of the personas, with a maximum value of 0.84.
Specifically, the personas 2, 4 and 8 scored highest in AUC, and also in Precision and Recall, whereas persona 6 has poor performance compared to the rest.
These results indicate that for high scoring personas, we successfully captured the active \name\ campaigns, but for the personas with lower scores, there may not be active campaigns for the period of the experiments.

In order to retrieve the variables that affect the discovery and measurement of \name, we applied the feature importance method on the dataset of each persona, and selected the top-10 highest scoring features.
For the majority of the personas (7 out of 10) the most important features were the number of ads (distinct or not) and the number of keywords in desktop. 
In some cases, there were also landing pages that had high scoring (i.e., specific ad-campaigns), but this was not consistent across all personas.

\noindent
\textbf{Combined Personas: Setup 1b.}
Here, we use all the datasets collected individually, for each persona in the previous experiment (Setup 1a), and combine them into one unified dataset.
This setup emulates the real scenario of a user exhibiting multiple and diverse web interests, that give extra information to the ad-ecosystem about their browsing behavior.
Of course, there is an increase in the possible feature space to accommodate all the domains and keywords from all personas.
In fact, the dataset contains 2021 features as it stores the vectors of landing pages and keywords, for all the different types of personas.
In total, there were 890 distinct ad-domains described by keywords in 76 distinct categories.

In this dataset, we apply feature selection with the Extra-Trees classifier to select the most relevant features and create a more accurate predictive model. 
This method reduced the feature space to 984 useful features out of 2201. 
Next, we use the three classification algorithms and a range of hyper-parameters for each one. 
Also, we apply a 10-fold nested cross-validation method for selecting the best model (in terms of scoring performance) that can give us an accurate, non overly-optimistic estimation~\cite{cawley2010over}.
Again, the best selected model was Random Forest, with 200 estimators (trees) and 200 depth of each tree, with AUC=0.89 (bottom row in Table~\ref{tab:persona_eval_shorttrain}). 
The model's performance is high in all the mentioned scores, which indicates that the more diverse data the advertisers collect, the easier it is to identify the different user's devices.
This result is in line with Zimmeck et al.~\cite{zimmeck2017}, who attempted a threshold-based approach for probabilistic \name\ detection on real users' data, lending credence to our proposed platform's performance.

\begin{figure}[t]
    \centering
    \includegraphics[width=0.9\columnwidth]{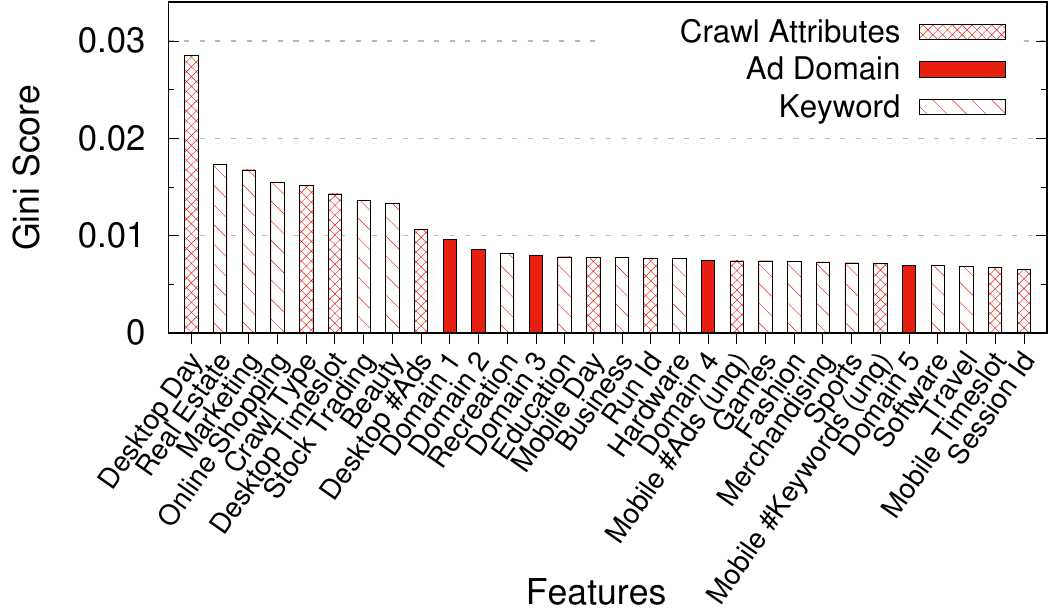}
    
    \caption{Top-30 features ranked by importance using Gini index, in the machine learning model.}

    \label{fig:importance}

\end{figure}

We also measure the feature importance for the top-30 features (shown in  Figure~\ref{fig:importance}).
One third of the top features are related to crawl specific metadata, whereas about half of the top features are keyword-related.
Interestingly, features such as the day and time of the experiment, as well as the number of received ads, are important for the algorithm to make the classification of the devices.
Indeed, time-related features provide hints on when the ad-ecosystem receives the browsing signal and attempts the \name, and thus, which days and hours of day the \name\ is stronger.
These results give support to our initial decision to experiment in a continuous fashion with regular sessions injecting browsing signal, while at the same time measuring the output signal via delivered ads.

\subsection{Does long-time browsing improve \name?}\label{sec:longtrain} 

\noindent
\textbf{Independent Personas: Setup 2a.}
In this set of experiments, we allow the devices to train for a longer period of time, to emulate the scenario where a user is focused on a particular interest, and produces heavy browsing behavior around a specific category.
This long-lived browsing injects a significantly higher input signal to the ad-ecosystem than the previous setup, which should make it easier to perform \name.
In order to increase the setup's complexity, and make it more difficult to track the user, we allow all devices (i.e., 1 mobile, 2 desktops) to train in the same way under the same persona.
In effect, this setup also tests a basic countermeasure from the user's point of view, who tries to blur her browsing by injecting traffic of the same persona from all devices to the ad-ecosystem.

In this setup, while all devices are trained with the same behavioral profile, we examine if the statistical tests and ML modeler can still detect and distinguish the \name.
This experiment contains three different phases during each run.
The mobile phase, where the mobile performs training crawls for $t_{train}$=480 mins, and a testing crawl for $t_{test}$=30 mins.
In parallel with the mobile training, the two desktops perform test crawls for $t_{test}$=30 mins.
After mobile training and testing, both desktops start continuous training and testing crawls alternately for 8 hours ($t_{train}$=$t_{test}$=30 min).

Due to the long time needed for executing this experiment, we focus on two personas constructed in the following way.
We use the methodology for persona creation as described in~\cref{sec:input}, and focus on active ad-campaigns, resulting to two personas in the interest of ``Online Shopping-Accessories'', and ``Online Shopping-Health and Fitness'' (loosely matching the personas 1 and 4 from Table~\ref{table:personas}).
Then, we performed 4 runs of 16 hours duration each, for each persona.
In this setup, since all devices are uniformly trained, we do not include the keyword vector of the persona pages into the datasets, to not introduce any bias from repetitive features.

The statistical analysis for this experiment reveals potential \name, since we accept the null hypothesis for the distribution of ads delivered in the paired desktop and mobile (lowest p-value=0.052), and reject it in the baseline desktop and mobile (highest p-value=0.006).
This consistency is interesting, since for this setup all three devices are uniformly trained with the same persona, and thus all of them collect similar ads due to retargeting.
However, there is no similarity between the distributions of ads in the devices that do not share the same IP address.

To clarify this finding, we applied the ML algorithms as in the previous experiment.
The algorithms again detect \name\ between the mobile and the paired desktop, even though all devices were exposed to similar training with the same persona.
In fact, Logistic Regression performed the best across both personas, with AUC $\ge$ 0.81, and $F_1$-score $\ge$ 0.80 for both classes.Detailed evaluation results of ~\cref{sec:longtrain} presented in  able~\ref{appendix:persona_eval_2}.
When computing the importance of features, the desktop number of ads and keywords and the desktop time slot are in the top-10 features.
Based on these observations, we believe that the longer training time allowed the ad-ecosystem to establish an accurate user profile, and retarget ads on the paired desktop, based on the mobile's activity.

\begin{table}[t!]
	\centering
	\caption{Performance evaluation for Logistic Regression in total components of  Setup 2.
		Left value in each column is the score for Class 0 (C0=not paired desktop); right value for Class 1 (C1=paired desktop).}
	\label{appendix:persona_eval_2}
	\small
	\resizebox{0.85\columnwidth}{!}{
	\begin{tabular}{c|cc|cc|cc|c}
		\toprule
		{\textbf{Persona}}&\multicolumn{2}{c|}{\textbf{Precision}}&\multicolumn{2}{c|}{\textbf{Recall}}&\multicolumn{2}{c|}{\textbf{$F_1$-Score}}&\multirow{2}{*}{\textbf{AUC}}\\
		\cline{2-3}\cline{4-5}\cline{6-7}
		(setup) & \textbf{C0}&\textbf{C1} & \textbf{C0}&\textbf{C1}	& \textbf{C0}&\textbf{C1} &\\		
		\toprule
		1  (2a) & 0.90 & 0.79 & 0.82 & 0.88 & 0.86 & 0.83 &\textbf{0.85}\\
		
		4 (2a) & 0.83 & 0.79 & 0.81 & 0.81 & 0.82 & 0.80 &\textbf{0.81}\\
		
		combined(2b) & 0.87 & 0.92 & 0.92 & 0.87 & 0.89 & 0.90 &\textbf{0.89}\\
		\hline
		\rowcolor{Gray}
		
		1 (2c) & 0.87 & 1.0 & 1.0 & 0.88 & 0.93 & 0.93 &\textbf{0.93}\\
		\rowcolor{Gray}
		
		4 (2c) & 1.0 & 0.98 & 0.98 & 1.0 & 0.99 & 0.99 &\textbf{0.99}\\
		\rowcolor{Gray}
		
		combined(2d) & 1.0 & 0.86 & 0.88 & 1.0 & 0.93 & 0.93 &\textbf{0.93}\\
		\bottomrule
	\end{tabular}
		}
\end{table}

\noindent
\textbf{Combined Personas: Setup 2b.}
Similarly to \cref{sec:shorttrain} we combine all data collected from the Setup 2a into a unified dataset.
Under this scenario, in which we mix data from both personas, the classifier again performs well, with AUC=0.89.
Important features in this case are the number of ads and keywords delivered to the desktops, the time of the experiment, and number of keywords for the desktop.

\noindent
\textbf{Boosted Browsing with \name\ trackers and Independent Personas: Setup 2c.}
In the next set of experiments, we investigate the role of \name\ trackers in the discovery and measurement of \name.
In particular, we attempt to boost the \name\ signal, by visiting webpages with higher portion of \name\ trackers.
Therefore, the experimental setup and the preprocessing method remain the same as in the previous Setup 2a, but 
we select webpages to be visited that have active ad-campaigns and their landing pages embed the most-known \name\ trackers (as we also show in the next section): Criteo, Tapad, Demdex, Drawbridge.
We also change the set of our control pages, so that each one contains at least a \name\ tracker.
News sites have many 3rd-parties compared to other types of sites~\cite{englehardt_ccs}.
Thus, for this boosted browsing experiment, we choose the set of control pages to contain 3 weather pages and 2 news websites,\footnote{\url{accuweather.com}, \url{wunderground.com}, \url{weather.com}, \url{usatoday.com}, \url{huffingtonpost.com}} while verifying they do not serve contextual ads.

Performing the same analysis as earlier, we find that mobile and paired desktop have ads coming from the same distribution (lowest p-value=0.10), and that there is no similarity between the ads delivered in the mobile and baseline desktop (highest p-value=0.007).
For a clearer investigation of the importance of the \name\ trackers, we also evaluate the findings with the ML models.
For persona 1, Logistic Regression and Random Forest models perform near optimally, with high precision of Class 1, high recall for class 0, average $F_1$-Score=0.93 for both classes, and AUC=0.93.
For persona 4, the scores are even higher, outperforming the other setups, as all metrics for Logistic Regression scored higher than 0.98.
Overall, these results indicate that we successfully biased the trackers to identify the emulated user in both devices, and to provide enough output signal (ads delivered) for the statistical algorithms to detect the \name\ performed.

\noindent
\textbf{Boosted Browsing with CDT trackers and Combined Personas: Setup 2d.}
We follow a similar approach with before, and combine all data collected from the Setup 2c, into a unified dataset for Setup 2d.
Under this scenario, the classifier (Logistic Regression) again performs very well, with AUC=0.93.
Important features in this case are the number of ads delivered to the desktops, the time of the experiment in each desktop and the number of keywords.
Interestingly, and perhaps unexpectedly, the existence of Criteo tracker in a landing page, is a feature appearing in the top-10 features.

\subsection{Does incognito browsing help evade \name?}\label{sec:stateless}
\noindent
\textbf{Independent Personas: Setup 3a.}
In this final experimental setup, we investigate if it is possible for the user to apply some basic countermeasures to avoid, or at least reduce the possibility of \name, by removing her browsing state in every new session.
For this, we perform experiments where the traditional tracking mechanisms (e.g., cookies, cache, browsing history, etc.) are disabled or removed, emulating incognito browsing.
We select the first five personas from Table~\ref{table:personas}, which had the most active ad-campaigns and appeared to be promising due to the ``online shopping'' interest.
Every desktop executed browsing in a stateless mode, while the mobile in a stateful mode.
For each persona, we collected data for two runs, following the timeline of phases as in Setup 1a.

The distributions between mobile vs. paired desktop, as well as mobile vs. baseline desktop, were found to be different (highest p-value=0.034).
Also, none of the ML classifiers performed higher than 0.7 (in all metrics), and thus we could not clearly extract any significant result.
Specifically, the highest AUC score for personas 1 and 2 was 0.70 with the use of the Random Forest classifier, and for personas 3 and 4 was 0.73 using the Logistic Regression classifier.
The worst scoring, independent of algorithm, was recorded for persona 5, with AUC=0.57, and Precision/Recall scores under 0.50.

\noindent
\textbf{Combined Personas: Setup 3b.}
When the data from all five personas are combined, the classifier performing best was Logistic Regression, with AUC=0.79.
Overall, these results point to the semi-effectiveness of the incognito browsing to limit \name.
That is, by removing the browsing state of a user on a given device, the signal provided to the \name\ entities is reduced, but not fully removed.
In fact, when the data from various personas are combined, the \name\ is still somewhat effective, since the paired devices have the same IP address.

\section{Platform Validation}
\label{sec:platformval}

In this section we validate the representativeness of the data collected from the previous experiments, by examining: (i) the type and frequency of ads delivered in each device, and (ii) the type and number of trackers that our personas were exposed to.
We compare the distributions of these quantities with past works and data on real users, to quantify if our synthetic personas successfully emulate real users' traffic, and if our measurements of the \name\ ad-ecosystem are realistic.

\begin{figure}[t]
	\centering
	\includegraphics[width=0.9\columnwidth]{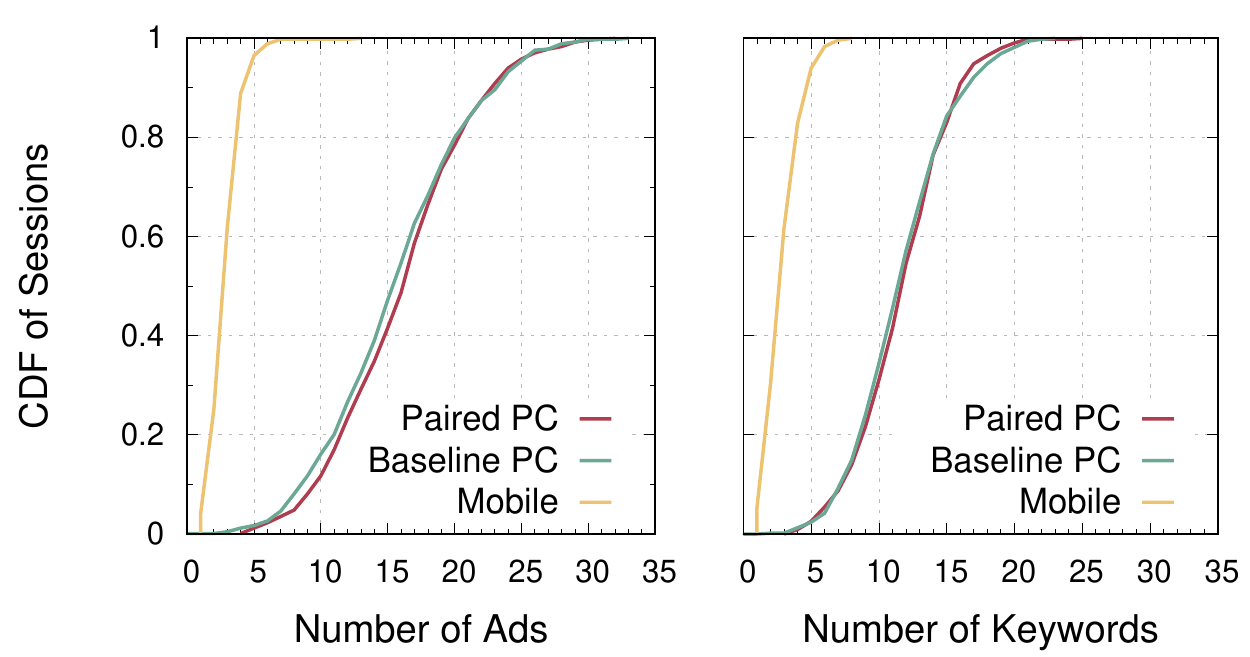}
	\caption{CDF of collected ads (left) and corresponding keywords of the ads (right) per crawling session for all devices.}
	\vspace*{-4mm}
	\label{fig:cdf_ads_keys}

\end{figure}

We first measure the frequency of ads delivered to our devices in the experiment~\cref{sec:shorttrain}, since it follows a  well-crafted timeline that is suitable for this kind of measurement.
The ads delivered in the  three devices during these sessions are shown in Figure~\ref{fig:cdf_ads_keys} (left).
For most sessions ($\sim$90\%), the mobile device was exposed to fewer than five ads, since the mobile version of websites typically delivers a smaller number of ads, designed for smaller screens and devices.
On the contrary, the desktop devices had a higher exposure to ads compared to the mobile device. 
Also, the two desktops receive a similar number of ads (on average 2 to 4 ads on every visit to the control pages).
Similar observations can be made for the keywords categories of ads  (Figure~\ref{fig:cdf_ads_keys} (right)).
The ad-industry has reported that $\sim$300 ads everyday, on average, are being displayed to desktop users~\cite{forbs_ads,biz_ads,tsc_ads,huff_ads}, while  they also recommend the delivery of 5 ads per mobile domain~\cite{mobile_limit}, which proportionally match the number of ads we have collected  in our mobile and desktop sessions.

\begin{figure}[t]
	\centering
	\includegraphics[width=0.9\columnwidth]{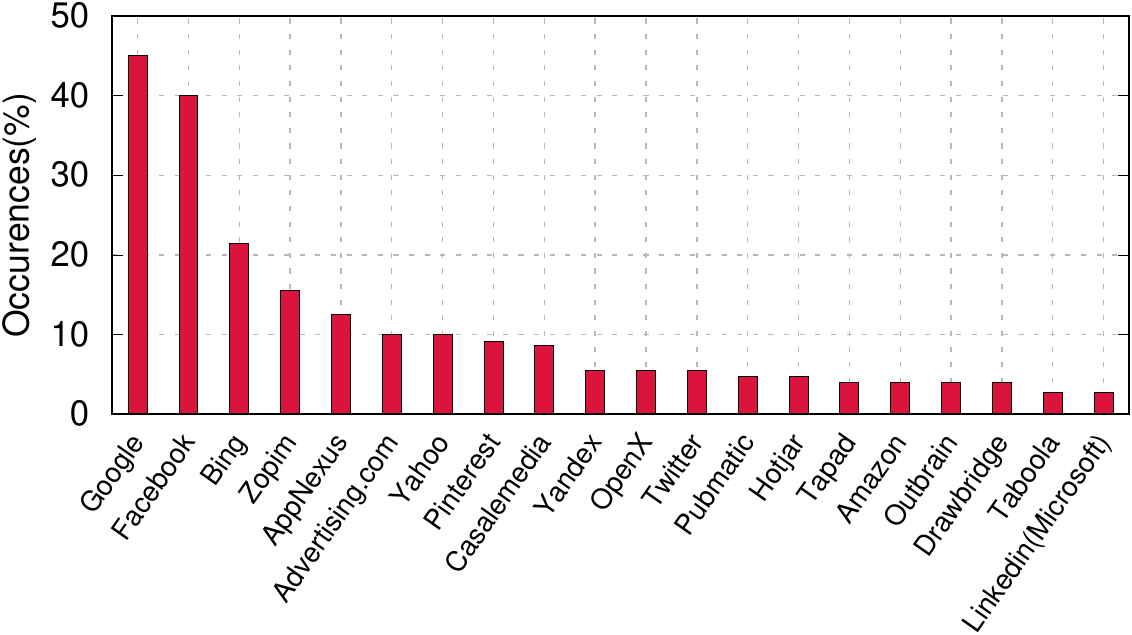}
	\caption{Top-20 trackers (grouped based on organization) and their coverage in persona pages. For example, all the Google-owned domains, such as Doubleclick, Googleapis, Google-Analytics, are grouped under the ``Google'' label.
	}
	\label{trackers_freq}
\end{figure}

We also validate the representativeness of the data collected from the experiments~\cref{sec:shorttrain} and \cref{sec:longtrain}, by  examining the trackers appearing in the webpages visited by the personas.
We use Disconnect List~\cite{disco} to detect them and measure their frequency of appearance (i.e., Figure~\ref{trackers_freq}).
From the trackers detected in the set of persona pages, and using the list provided by~\cite{zimmeck2017}, $37\%$ was found to be \name\ related, including both deterministic and probabilistic.
In fact, the top \name\ trackers found in our data, which may perform both types of \name, include Google-owned domains,
Facebook, Criteo, Zopim, Bing, Advertising.com(AOL), and are in-line with the top \name\ trackers found in~\cite{zimmeck2017, brookman2017} (66\% overlap of top-20 with~\cite{zimmeck2017} and 55\% overlap with~\cite{brookman2017}).
In addition, $17\%$ of these trackers are mainly focused on probabilistic \name, including Criteo, BlueKai, AdRoll, Cardlytics, Drawbridge, Tapad, and each individual tracker is found at least in $2\%$ of the persona pages, again in-line with the results in~\cite{zimmeck2017}.


\section{Discussion \& Conclusion}
\label{sec:discussion}

Through extensive experiments with the proposed framework \fw, we were able to trigger \name\ trackers into pairing of the emulated users' devices.
This allowed us to statistically verify that \name\ is indeed happening, and measure its effectiveness on different user interests and browsing behaviors, independently and in combination.
In fact, \name\ was prominent when user devices were trained to browse pages of similar interests, reinforcing the behavioral signal sent to \name\ entities, and specifically when browsing activity is related with online shopping, since those types of users seem to be more targeted by advertisers.
The \name\ effect was further amplified when the visited persona and control pages had embedded \name\ trackers, pushing the accuracy of detection up to 99\%.
We also found that browsing in a stateless mode showed a reduced, but not completely removed \name\ effect, as incognito browsing obfuscates somewhat the signal sent to the ad-ecosystem, but not the network access information.
Indeed, our data collection was performed across relatively short time periods, in comparison to the wealth of browsing data that advertising networks have at their disposal.
In fact, we anticipate that \name\ companies collect data about users and devices 
for months or years, and even buy data from data brokers, to have the capacity of targeting users with even higher rates.
To that end, we believe that high accuracies self-reported by \name\ companies (e.g., Lotame: $>$90\%~\cite{lotame}, Drawbridge: 97.3\%~\cite{draw_acc}), are possible.\\
\noindent
\textbf{Impact on user privacy:}
Undoubtedly, \name\ infringes on users' online privacy and minimizes their anonymity.
But the actual extent of this tracking paradigm and its consequences to users, the community, and even to the ad-ecosystem itself, are still unknown.
In fact, since \name\ is heavily depended on user's browsing activity, and the ad-ecosystem employs such collected data for targeting purposes, one major line of future work is the study of targeting sensitive user categories (e.g., gender, sexual orientation, race, etc.) via \name.
This is especially relevant nowadays with the enforcement of recent EU privacy regulations such as GDPR~\cite{eu:gdpr} and ePrivacy~\cite{eprivacy}.
This is where \fw\ comes in play, as it provides a concrete, scalable and extensible methodology for experimenting with different \name\ scenarios, auditing its mechanics and measuring its impact.
In fact, the modular design of our methodology allows to study \name\ in depth, and propose new extensions to study the \name\ ecosystem: new plugins, personas and ML techniques.
To that end, our design constitutes \fw\ into an enhanced transparency tool that reveals potentially illegal biases or discrimination from the ad-ecosystem.



\section*{Acknowledgments}

The research leading to these results has received funding from the European Union's Horizon 2020 Research and Innovation Programme under grand agreement No 786669 
(project CONCORDIA), the Marie Sklodowska-Curie grant agreement No 690972 (project PROTASIS), and the Defense Advanced Research Projects Agency (DARPA) ASED Program and AFRL under contract FA8650-18-C-7880. The paper reflects only the authors' views 
and the Agency and the Commission are not responsible for any use that may be made of the information it contains.

{\normalsize \bibliographystyle{acm}
	\bibliography{paper.bib}}

\end{document}